\documentclass[a4paper,11pt]{article}
\usepackage{aaskaiid}
\usepackage{orcidlink}

\usepackage{aas_macros}
\usepackage{url}

\newcommand{\hii}{H{\sc ii}}
\newcommand{\degree}{$^{\circ}$}

\usepackage[normalem]{ulem}

\title{Probing Anomalous Microwave Emission with the Square Kilometre Array}
\ShortTitle{AME with SKA}

\author[1]{Mat\'ias~Vidal\orcidlink{0000-0003-2134-1587}}
\ShortName{Vidal et al.} 
\author[2,3]{Sim\'on~Casassus\orcidlink{0000-0002-0433-9840}}
\author[4,5]{Roke~Cepeda-Arroita\orcidlink{0000-0002-9043-2645}}
\author[6,3]{Miguel~C\'arcamo\orcidlink{0000-0003-0564-8167}}
\author[7]{Stuart~E.~Harper\orcidlink{0000-0001-7911-5553}}
\author[8,9]{Thiem~Hoang\orcidlink{0000-0003-2017-0982}}
\author[4,5]{J.~A.~Rubiño-Martín\orcidlink{0000-0001-5289-3021}}
\author[10]{Kieran~A.~Cleary\orcidlink{0000-0002-8214-8265}}
\author[7]{Clive~Dickinson\orcidlink{0000-0002-0045-442X}}
\author[4,5]{R.~T.~G\'enova-Santos\orcidlink{0000-0001-5479-0034}}
\author[7]{Gabriel~A.~Hoerning\orcidlink{0000-0002-8677-6656}}
\author[11]{Melis~O.~Irfan\orcidlink{0000-0003-2021-7357}}
\author[12]{A.~Lazarian\orcidlink{0000-0002-7336-6674}}
\author[13,14]{Hauyu~Baobab~Liu\orcidlink{0000-0003-2300-2626}}
\author[15]{Eric~J.~Murphy\orcidlink{0000-0001-7089-7325}}
\author[16]{M.~W.~Peel\orcidlink{0000-0003-3412-2586}}
\author[17,18]{Nathalie~Ysard\orcidlink{0000-0003-1037-4121}}
\author[7]{Zheng~Zhang\orcidlink{0000-0002-9154-2803}}

\affiliation[1]{Universidad Aut\'onoma de Chile, Facultad de
  Ingenier\'ia, N\'ucleo de Astroqu\'imica \& Astrof\'isica. Av. Pedro
  de Valdivia 425, Providencia, Santiago, Chile.}
\affiliation[2]{Departamento de Astronomía, Universidad de Chile,
  Casilla 36-D, Santiago, Chile}
\affiliation[3]{Data Observatory Foundation, Eliodoro Yáñez 2990,
  Providencia, Santiago, Chile}
\affiliation[4]{Instituto de Astrofísica de Canarias, E-38200 La Laguna, Tenerife, Spain}
\affiliation[5]{Departamento de Astrofísica, Universidad de La Laguna, E-38206 La Laguna, Tenerife, Spain}
\affiliation[6]{University of Santiago of Chile (USACH), Faculty of
  Engineering, Computer Engineering Department, Chile}
\affiliation[7]{Jodrell Bank Centre for Astrophysics, Department of Physics and Astronomy, School of Natural Sciences, The University of Manchester, Alan Turing building, Oxford Road, Manchester M13 9PL, UK}
\affiliation[8]{Korea Astronomy and Space Science Institute, Daejeon 34055, Republic of Korea}
\affiliation[9]{Department of Astronomy and Space Science, University of Science and Technology, 217 Gajeong-ro, Yuseong-gu, Daejeon, 34113,
  Republic of Korea}
\affiliation[10]{California Institute of Technology, 1200 E. California Blvd., Pasadena, CA 91125, USA}
\affiliation[11]{Institute of Astronomy, University of Cambridge, Cambridge, CB3 0HA, UK}
\affiliation[12]{Department of Astronomy, University of Wisconsin-Madison, Madison, WI 53706, USA}
\affiliation[13]{Department of Physics, National Sun Yat-Sen University,
No. 70, Lien-Hai Road, Kaohsiung City 80424, Taiwan, R.O.C.}
\affiliation[14]{Center of Astronomy and Gravitation, National Taiwan Normal University, Taipei 116, Taiwan}
\affiliation[15]{National Radio Astronomy Observatory, 520 Edgemont Road, Charlottesville, VA 22903, USA}
\affiliation[16]{Imperial College London, Blackett Lab, Prince Consort Road, London SW7 2AZ, UK}
\affiliation[17]{Institut de Recherche en Astrophysique et Planétologie, Université Toulouse III - Paul Sabatier, CNRS, CNES, 9 Av. du colonel
Roche, 31028 Toulouse, France}
\affiliation[18]{Université Paris-Saclay, CNRS, Institut d’Astrophysique Spatiale, 91405 Orsay, France}

\emailAdd{matias.vidal@uautonoma.cl}

\abstract{

  Anomalous microwave emission (AME) represents an excess of radiation in the 10--60\,GHz range, distinct from synchrotron, free-free, or thermal dust emission.  Although most commonly attributed to electric dipole radiation from rapidly rotating small dust grains (spinning dust), alternative mechanisms such as magnetic dipole emission (MDE) remain plausible. The detection of AME across diverse environments, from diffuse interstellar clouds to protoplanetary disks and external galaxies, suggests that multiple physical processes or carriers may contribute to its origin. Understanding AME is essential for both Galactic astrophysics and cosmology, as it constitutes a significant foreground for cosmic microwave background (CMB) studies, potentially biasing measurements.

  This chapter reviews current theoretical frameworks and observational evidence for AME, highlighting the key outstanding questions concerning its emission mechanisms, carriers, and polarization properties. We discuss how the Square Kilometre Array Observatory (SKAO), through its unprecedented sensitivity, angular resolution, and frequency coverage, will transform AME studies. SKA observations will enable detailed mapping of AME morphology, precise characterisation of its spectral energy distribution, and the identification of its carriers in Galactic and extragalactic environments. By combining SKA-mid data with higher-frequency observations from ALMA and other facilities such as SPHEREx, it will be possible to disentangle competing models and exploit AME as a diagnostic probe of interstellar grain physics and the small-scale structure of the interstellar medium.

}

\begin{document}
\maketitle

\section{Introduction \& Context}

Anomalous microwave emission (AME) refers to an excess signal observed in the $\sim$10--60\,GHz frequency range, beyond what is expected from the cosmic microwave background (CMB), and from free-free, synchrotron, and thermal dust emission from the interstellar medium in our galaxy and others. It was first identified in 1996 as a foreground contaminant in CMB studies \citep{kogut96b,Leitch1997}, where its presence complicates the accurate modelling and separation of foregrounds, which remains an important challenge for both current and future CMB experiments (e.g., \citealt{Planck_Bicep2:2015,Planck2016_polDust}). AME has been detected in a wide variety of Galactic environments: as diffuse emission at high Galactic latitudes \citep{kogut96b,Leitch1997,deOliveira-Costa2004,Finkbeiner2004b}, in molecular clouds and near \hii\ regions \citep{Watson2005,Davies2006,Casassus2008,Vidal2011,Dickinson2013}, in a few proto-planetary disks \citep{Greaves2018}, and in extragalactic sources \citep[e.g.][]{Murphy2010,Murphy2018,Linden2020}.

Several models have been proposed to explain the AME. The most widely accepted is the spinning dust hypothesis (see Section~\ref{sec:SpinningDust}), in which very small dust grains emit electric dipole radiation as they rotate at GHz frequencies. This model successfully reproduces the concave spectra observed in many regions by the {\em Planck} satellite using realistic physical parameters for interstellar clouds \citep{planck_sd:11,Planck_SD2014}. However, it does not fully explain all observations, such as some extragalactic detections (see Section \ref{sec:extragalactic}). An alternative mechanism is magnetic dust emission, which does not rely on grain rotation but instead arises from thermal fluctuations in the magnetisation of magnetic nanoparticles (see Section~\ref{sec:MagneticDust}).

The detection of AME in such a diverse range of environments remains puzzling. It is not yet clear whether there is a common factor among these sources, related either to the dust grain properties (for example, size distribution or composition) or to the local environmental conditions (such as radiation field, density, or temperature), that promotes the emergence of AME. One of the few robust observational patterns is that AME is consistently associated with photo-dissociation regions (PDRs) \citep{Dickinson2018}, which suggests that the specific physical and chemical conditions in PDRs play a key role in enabling or enhancing this emission.

To understand this emission mechanism, precise observations are required across a broad frequency range. AME typically peaks within 10--60\,GHz, where other diffuse mechanisms such as free-free and synchrotron emission also contribute. The free-free and synchrotron components can be constrained using data at frequencies $\lesssim 10$\,GHz. Thermal dust emission may also contribute within the AME range, which makes high-frequency ($\gtrsim 100$\,GHz) observations necessary to constrain this component. Good angular resolution is also important, since it allows the spatial separation of the different emission mechanisms. Because AME is usually associated with PDRs and \hii\ regions, free-free emission is often adjacent and must be carefully disentangled. In addition, detailed correlations with mid-infrared templates tracing small dust grains require angular resolutions of a few arcseconds in order to exploit the full potential of the mid-infrared data.

The spinning dust and magnetic dust mechanisms depend strongly on the nature of the emitters as well as on the physical conditions of the interstellar medium (ISM). Therefore, precise observations of AME can help identify the emitting particles and diagnose the ISM conditions, providing a powerful new tool for studying the interstellar environment.

This work expands on the AME science case presented in \citet{Dickinson2015}. The rest of this chapter begins by outlining the main open questions regarding AME, including the identification of the emission mechanisms and possible carriers (Section~\ref{sec:1-questions}). In Section~\ref{sec:2-observations}, we summarise the current observational status. Section~\ref{sec:3-SKA} discusses the role that the SKAO will play in advancing AME studies. Finally, Section~\ref{sec:4-conclusions} presents the conclusions.

\section{Outstanding Questions in AME}
\label{sec:1-questions}
The fact that AME is observed in such a diverse range of environments
is puzzling. Is there a common factor among these sources, whether
related to the dust grain properties (e.g., size distribution,
chemistry) or the local environment (e.g., radiation field, density,
temperature), that promotes the emergence of AME?

Here, in Section \ref{sec:EmissionMechanisms} we will discuss the main
candidates for the emission mechanisms, in Sec. \ref{sec:Carriers}
what are the most plausible carriers, in Sec. \ref{sec:Spectrum}
explore the possibility of measuring structure in the AME spectrum, in
Sec. \ref{sec:Environment} how does AME relates to environmental
conditions, and finally in Sec. \ref{sec:Polarisation}, explore the
polarisation properties of AME.

\subsection{What is the Dominant Emission Mechanism?}
\label{sec:EmissionMechanisms}

A variety of explanations have been proposed for the AME, such as
flat-spectrum synchrotron radiation \citep{Bennett2003} and hot
($T_\mathrm{e} \sim 10^6$~K) free-free emission
\citep{Leitch1997}. For a more detailed review on mechanisms, see
\citet{Dickinson2018}. At present, two main mechanisms are considered
plausible: electric dipole radiation from small, rapidly rotating dust
grains, and thermal fluctuations in magnetic dust particles. Although
the spinning-dust hypothesis remains the most widely accepted
explanation, the contribution of magnetic dust cannot yet be ruled
out. In the following, we briefly review the relevant theory and
predictions.

\subsubsection{Spinning Dust}
\label{sec:SpinningDust}
The leading explanation for AME is the spinning dust hypothesis, \citep{Erickson1957,DL1998a,DL1998b}, in which very small grains rotating at GHz frequencies emit electric dipole radiation. This model predicts a broad, bell-shaped spectrum peaking between $\sim$\,10\,--\,100\,GHz, depending on the physical conditions of the environment and the properties of the dust grains. A dust grain with electric or magnetic dipole moment $\mu$ rotating at angular frequency $\omega$ radiates according to the Larmor formula a power $P \propto \mu^{2}\,\omega^{4}$ at frequency $\nu = \omega/2\pi$, so the emission spectrum is set by the distributions of grain rotation rates and dipole moments. For a spherical grain of radius $a$ and density $\rho$ in thermal equilibrium with gas at temperature $T$, the characteristic frequency is

\begin{equation}
 \nu = 21\,{\rm GHz}\,\left(\frac{T}{100\,{\rm K}}\right)^{1/2}\,\left(\frac{\rho}{3\,{\rm g}\,{\rm cm}^{-3}}\right)^{-1/2} \,\left(\frac{a}{0.5\,{\rm nm}}\right)^{-5/2}~,
\end{equation}

implying that emission in the 20--30\,GHz range arises from grains with radii $a \lesssim 1$\,nm.

\begin{figure}[!t]
 \includegraphics[width=1\textwidth,angle=0]{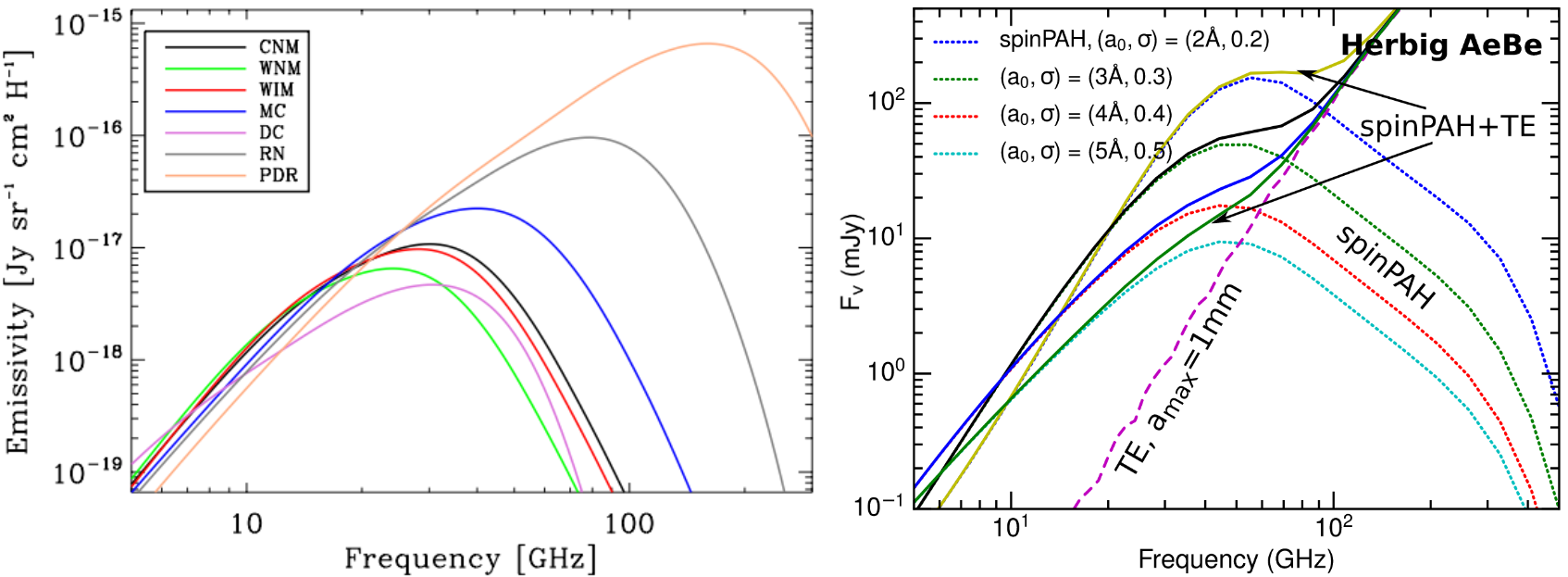}
 \caption{{\em Left:} Spinning dust emissivity spectra as a function of frequency for various idealised phases of the interstellar medium, as defined in \protect\cite{DL1998b}, computed using the {\sc SpDust2} code. The curves illustrate the strong dependence of emissivity on the local environmental conditions. {\em Right:} Spinning dust emissivity curves from spinning PAHs for different grain size distribution parameters. Adapted from \citet{Hoang2018}.}
 \label{fig:SD-emm}
\end{figure}

The observed spectrum of spinning dust emission arises from the sum of all individual grain spectra and thus depends directly on the distributions of rotation rates, dipole moments, and the alignments between dipoles and rotation axes. These distributions reflect the physical properties of the grains, such as their sizes, geometries, and compositions, which are themselves shaped by the local environment, including the intensity of the radiation field, gas temperature, and the abundances and ionisation states of hydrogen and carbon. Figure~\ref{fig:SD-emm} illustrates these dependencies: the {\em left} panel shows emissivity curves under varying environmental conditions, while the {\em right} panel highlights the effects of changing grain properties. Precise AME measurements thus provide a powerful probe of the interstellar medium and of the size distribution of the emitting grains, complementing information obtained at infrared wavelengths. Because AME is highly sensitive to local conditions, models of grain evolution, whether driven by ultraviolet photo-dissociation, shocks or adsorption, can be tested in a highly constraining manner.

\paragraph{Modelling of spinning dust}

By applying the Fokker--Planck formalism to compute the angular velocity distribution of interstellar dust grains, \texttt{spdust} \citep{Ali2009} enabled the calculation of the spinning dust spectral energy distribution (SED) through its publicly available IDL implementation. \citet{Hoang2010} refined the framework by introducing the wobbling motion of disk-like grains caused by internal thermal fluctuations and by accounting for transient spin-up using the Langevin equation. \citet{Silsbee2011} subsequently extended \texttt{spdust} to model the wobbling, oblate grains with random orientations in their updated \texttt{spdust2} code. More recently, \citet{Zhang25_Spydust} developed \texttt{SpyDust}, a Python implementation that generalises the treatment to allow arbitrary grain-shape distributions, while remaining fully compatible with the original \texttt{spdust2} framework.


\subsubsection{Magnetic Dust}
\label{sec:MagneticDust}

The gas-phase abundance of iron in the interstellar medium (ISM) is constrained primarily by ultraviolet scattering or absorption measurements of background starlight, which consistently show that Fe is one of the most heavily depleted elements in the neutral ISM \citep{Savage1996,Draine2003,Jenkins2009}. Although supernovae inject Fe into the gas phase \citep{Dwek2016}, measurements of Fe depletion indicate that along typical Milky Way sightlines $\ge$90\,\% of Fe is removed from the gas and locked up in dust \citep{Jenkins2009}, and in some slight-lines approaching close to 100\,\% depletion \citep{DeCia2021}. Plausible Fe materials in ISM include ferromagnetic oxides such as magnetite (Fe$_3$O$_4$) and maghemite ($\gamma$-Fe$_2$O$_3$), superparamagnetic Fe inclusions embedded within amorphous silicate grains, and a small population of free-flying metallic Fe nanoparticles.

In \citet{draine:99}, it was proposed that thermal fluctuations of the magnetisation in Fe materials would produce magnetic dipole emission (MDE) at microwave frequencies, offering an alternative complementary to spinning dust as an explanation for anomalous microwave emission (AME). In that framework the grain's magnetic response is modelled with a damped harmonic oscillator susceptibility, yielding a magnetic analogue of a Fr{\"o}hlich resonance in the absorption spectrum that typically peaks around $50$--$100$\,GHz (shape-dependent). It was later argued in \citet{Draine&Hensley2012} that MDE from magnetic nanoparticles could account for the sub-mm excess observed in some low-metallicity galaxies (e.g. the SMC), which motivated a refined treatment of MDE that incorporates a physically motivated Landau-Lifshitz-Gilbert susceptibility, explicit grain-shape effects, eddy currents, and updated predictions for spectra and polarisation \citep{DraineHensley2013}.

Unlike spinning-dust emission, MDE can exhibit high degrees of polarisation. The exact polarisation depends strongly on whether the magnetic material is present as free-flying single-domain grains or as magnetic inclusions embedded within larger amorphous silicate grains. For free-flying grains, models predict polarisation fractions up to tens of percent at microwave frequencies, with a pronounced frequency dependence and even possible flips in the polarisation angle at low frequencies that are dependent on grain shape \citep{draine:99,DraineHensley2013}. By contrast, for magnetic inclusions embedded in larger non-magnetic grains the expected polarisation is much smaller (typically at or below a few percent) \citep{draine:99}. Magnetically enhanced radiative torques can increase the alignment efficiency of the carrier grains when Fe inclusions are present, potentially pushing inclusion-dominated MDE to the percent-level regime, but still well below the free-flyer case \citep{Hoang&Lazarian2016}.

At present, observational evidence favours spinning dust as the dominant contributor to AME: the observed AME peak near $20$-$30$\,GHz is generally lower than the peak frequencies preferred by many MDE materials (often $\sim$50-100\,GHz for single-domain Fe or inclusion models), and current upper limits on AME polarisation are at the few percent level, disfavouring the highly polarised free-flyer case \citep{draine:99,DraineHensley2013}. That said, MDE is not ruled out; it may be subdominant in typical diffuse sightlines or become important in specific environments and frequency ranges. For example, as invoked to explain the sub-mm excess in the Small Magellanic Cloud \citep{Draine&Hensley2012}. Ultimately we expect to detect MDE at some level and putting constraints on MDE is critical for understanding the gas phase of the ISM, grain growth and destruction, and also as a polarised foreground to future polarised CMB missions. The properties of MDE can provide a valuable insight into the chemical composition and sizes of magnetic particles in the ISM.



\subsection{What are the Carriers?}
\label{sec:Carriers}

The nature of the AME carrier depends on the underlying emission
mechanism. Spinning dust emission may arise from multiple populations
of nanoparticles\footnote{Here we use the term nanoparticle (or
    nanograin) to refer to interstellar dust grains with
    characteristic sizes $\lesssim$100\,nm. However, spinning dust
    emission is dominated by the smallest members of this population,
    typically with sizes $\lesssim$10\,nm (often referred to as
    ultrasmall grains), owing to their higher rotational frequencies
    and larger electric dipole moments per unit mass.}, each with
different size, charge, and dipole moment distributions. Moreover,
especially in observations with low angular resolution, the detected
emission may arise from a combination of different interstellar
environments. These factors make it particularly challenging to
predict the expected spinning dust spectrum and to identify the
specific emitters directly from observations.In addition,

Carbonaceous and silicate nanograins are the leading candidates, as
they are expected to dominate the nanoparticle population in the
ISM. Other possibilities, such as nanodiamonds, have also been
proposed.

\subsubsection{PAHs / Carbon Nanograins}

Polycyclic aromatic hydrocarbons (PAHs) are widely accepted as the
carriers of the infrared features at 3.3, 6.2, 7.7, 8.6, and
12.7\,\textmu m \citep{Tielens2008}, arising from vibrational emission
following UV excitation \citep{LegerPuget1984}. They are pervasive in
the ISM, where they contribute to gas heating and chemistry.

Correlations between AME and PAH tracers have been reported in several
studies
\citep[e.g.][]{Casassus2006,Scaife2010,Ysard2010,Tibbs2011,Battistelli2015,Vidal2020,Casassus2021,Cepeda-Arroita2021}. However,
full-sky analyses using {\em Planck} and {\em WISE} data found little
or no correlation, suggesting that dust radiance may be a better AME
tracer \citep{Hensley2016,Chuss2022}.

This apparent tension has been revisited by \citet{Ysard2022}, who
accounted for gas column density and improved infrared data, showing
that carbonaceous nanograins remain consistent with the
observations. The remaining dispersion can be explained by variations
in grain size distribution, abundance, and dipole moments.

High-resolution radio and mid-infrared observations will be essential
to identify the emitting species. In particular, upcoming PAH maps
from {\em SPHEREx} \citep{Dore2014} will provide critical tests of
their role in AME.

\subsubsection{Nanosilicates}

A significant fraction of interstellar dust is composed of silicates
\citep{Henning2010}. In the ISM, grain-grain collisions driven by
supernova shocks can fragment $\sim$0.1\,\textmu m silicate grains
into nanosilicates, which are a leading candidate for the origin of
AME, although they have not yet been unambiguously detected.

Quantum chemical calculations indicate that nanosilicates possess
large intrinsic electric dipole moments, allowing even a small
fraction ($\sim$1\%) of the silicon dust mass to contribute
significantly to AME \citep{MaciaBromley2020}. Models by
\citet{Hoang2016} show that spinning nanosilicates can reproduce the
observed AME for log-normal size distributions, with constraints
$Y_{\mathrm{Si}}\lesssim10\%$ and $\beta\gtrsim0.4$\,D. Similar
conclusions were reached by \citet{Hensley2017}, consistent with
observational constraints from extinction and infrared emission.

More recent quantum chemical studies, however, find systematically
larger dipole moments ($\beta>1$\,D), which challenge these models
\citep{MaciaBromley2020,Marinoso2021}. Combined with full-sky analyses
\citep{Ysard2022}, this suggests that nanosilicates can contribute
only marginally to the AME unless their abundance is very low
($Y_{\mathrm{Si}}\sim1\%$).

\begin{figure}[!t]
 \centering
 \includegraphics[width=0.6\textwidth,angle=0]{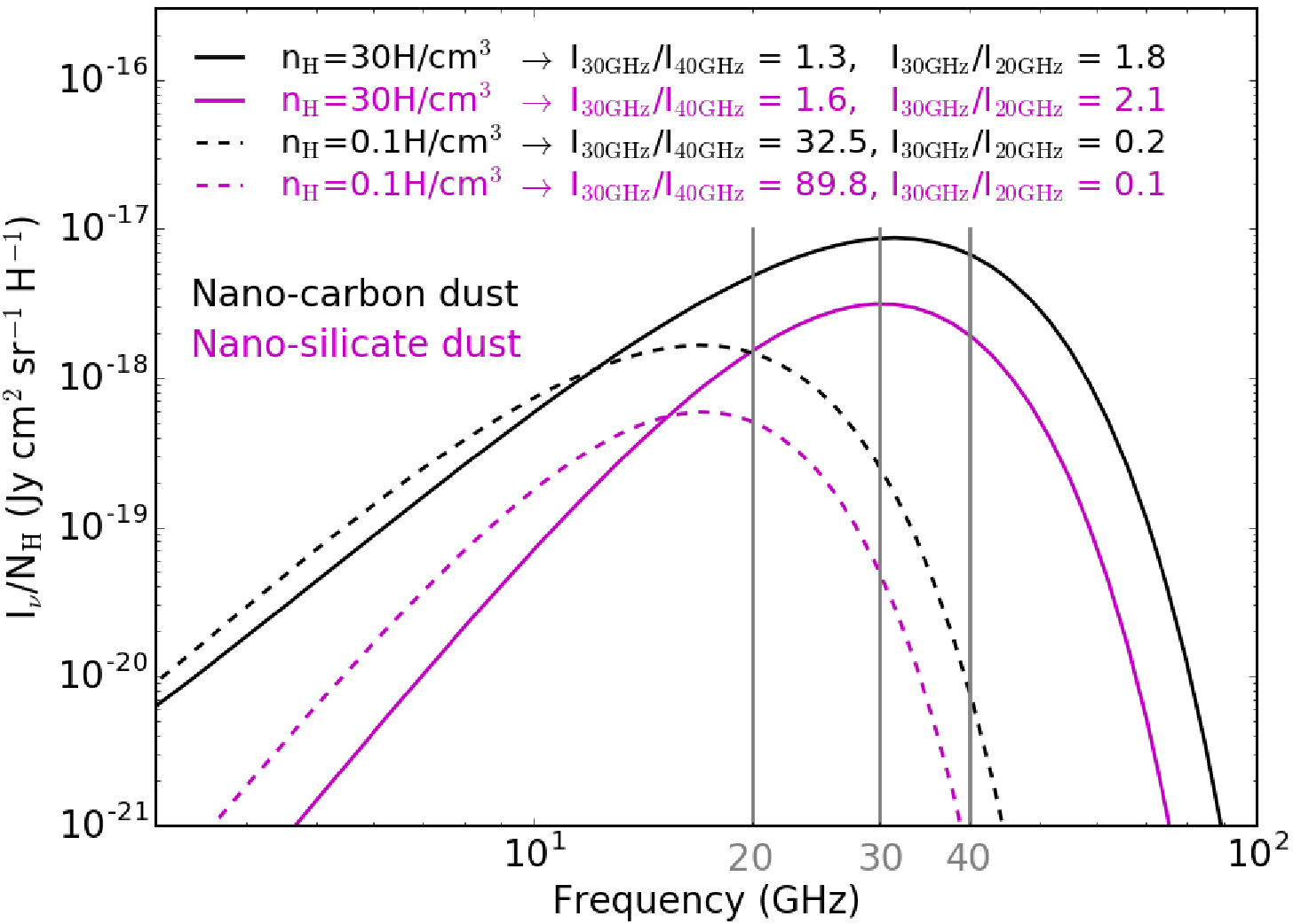}
 \caption{Spinning dust emission curves for nanocarbon and nanosilicate grains. To distinguish between them, it is necessary to measure over a wide frequency range, both below and above the typical AME peak at 30\,GHz. Adapted from \citet{Ysard2022}.}
 \label{fig:pah-silicates}
\end{figure}

A key discriminant between nanosilicates and nanocarbon grains is the
AME spectral energy distribution: nanosilicates are predicted to
produce narrower spectra on the low-frequency side. Observations below
20\,GHz are therefore critical for distinguishing between these
emission mechanisms (Fig.~\ref{fig:pah-silicates}).

\subsubsection{Nanodiamonds}

Hydrogenated nanodiamonds have been proposed as an alternative source of spinning emission. Identified in meteorites \citep{Lewis1987}, they are thought to produce the mid-infrared features at 3.43 and 3.53\,$\mu$m \citep{Guillois1999}. \citet{Greaves2018} reported AME detections in three protoplanetary discs hosting such particles; however, these results are based on single-dish data and show significant discrepancies with interferometric observations from the Karl G. Jansky Very Large Array (JVLA) and ALMA over the $4$--$400$\,GHz range \citep[][see Section~\ref{sec:compact}]{Chung2025}.



\subsection{Is There Structure in the AME Spectrum?}
\label{sec:Spectrum}

If anomalous microwave emission is indeed produced by spinning PAHs, the observed continuum spectrum would represent the collective rotational emission from individual molecules. Rotational spectroscopy in the $\sim$30--50\,GHz range of the spinning-dust emission can therefore be used to identify interstellar PAHs \citep{Ali-Haimoud2014}. Quasi-symmetric PAHs exhibit highly regular rotational spectra, appearing as a ``comb'' of evenly spaced stacks of lines. This type of molecule is believed to constitute a significant fraction of the interstellar PAH population \citep{Hudgins2005}, making them a promising target for rotational spectroscopy.

The first search for such lines using this method was carried out by \citet{Ali-Haimoud2015} with the Green Bank Telescope, targeting the Perseus molecular cloud, one of the brightest AME sources in the northern hemisphere. However, no signal was detected. Although the Perseus cloud was selected for its brightness, it may not be an ideal target, as a compact radio source lies close to the peak of its AME emission, with free-free radiation accounting for at least 30\% of the flux at 30\,GHz \citep{Tibbs2013}.

\citet{McGuire2018,mcguire_detection_2021} detected nitrogen-bearing PAHs in the Taurus Molecular Cloud (TMC) using the Green Bank Telescope. More recently, \citet{Wenzel2024} reported the detection of a four-ring PAH in the same cloud. These observations confirm the presence of such molecules in the interstellar medium and demonstrate the feasibility of detecting their rotational lines. Moreover, AME has been detected in the TMC by the QUIJOTE experiment \citep{Poidevin2019}, with an angular resolution of $1^\circ$. Given the coarse beam of QUIJOTE, no strong conclusion can yet be drawn, as the emission may arise from different regions within the cloud. The link between PAHs and AME therefore remains unconfirmed, although these results suggest that further investigations with improved angular resolution could help resolve this question. A detection of PAH rotational lines in a bright AME source would provide compelling confirmation of PAHs as the dominant emitters responsible for AME.

\begin{figure}[!t]
  \centering
 \includegraphics[width=0.7\textwidth,angle=0]{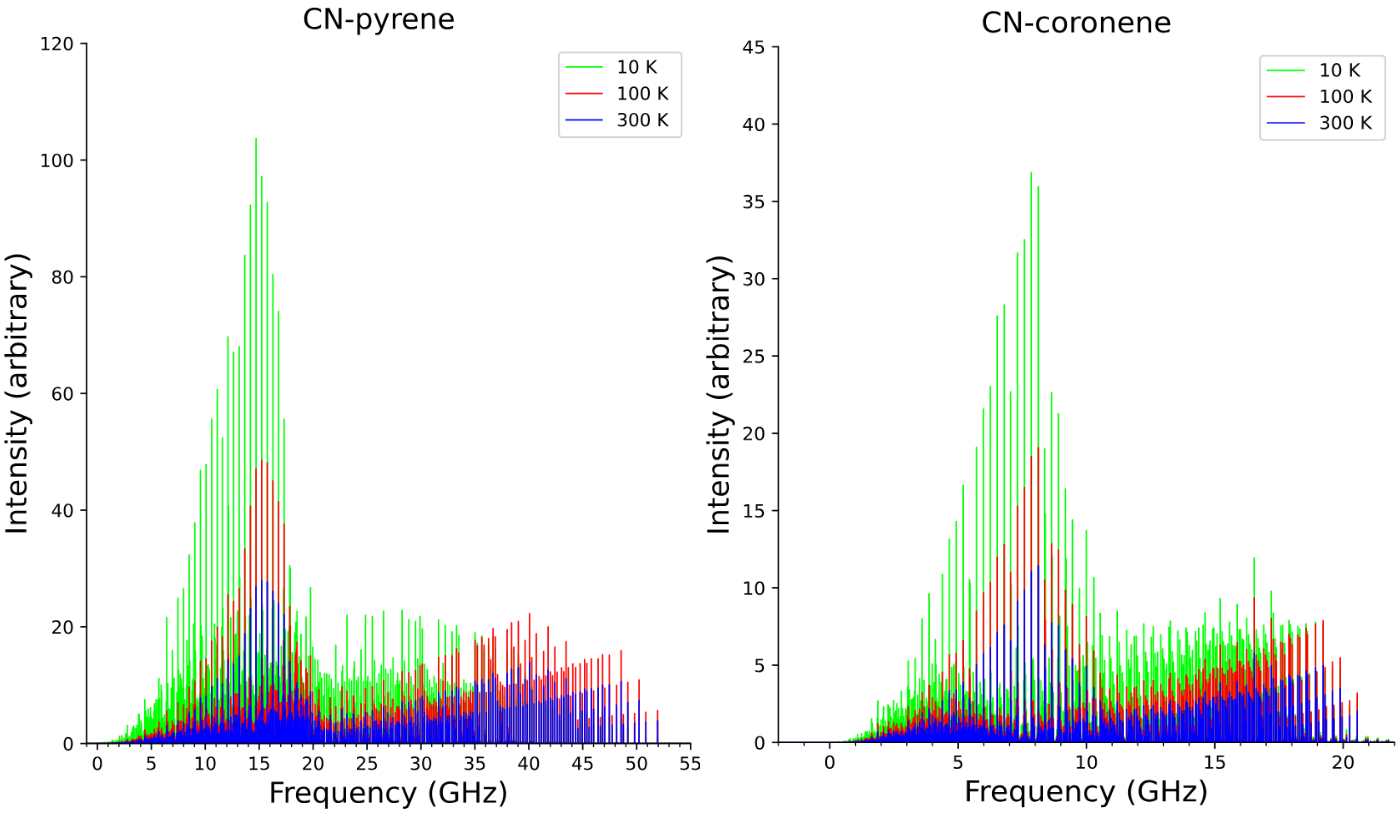}
 \caption{Calculated rotational spectra of CN-PAHs at 10 K (green), 100 K (red), and 300 K (blue). Figure adapted from \citet{Vats2022}.}
 \label{fig:CNPAH}
\end{figure}

Figure~\ref{fig:CNPAH} shows the calculated rotational spectra for two cyanide-substituted PAHs that could potentially be identified through observations at frequencies $\lesssim 15$\,GHz.

\subsection{How does AME change with local physical conditions?}
\label{sec:Environment}

Understanding how the AME changes with the physical properties of local conditions (gas, dust, and radiation field) is essential to establish AME as a diagnostic tool for the nanodust population.

Theoretical predictions show a weak dependence of AME from spinning dust on the local radiation field \citep{Ali2009}. Still, it depends strongly on the abundance of VSG and the gas temperature due to the dependence of the AME emissivity on the thermal rotation rate determined by gas collisions \citep{Hoang2010}. However, recent observations show that AME varies with the local radiation field (\citealt{Cepeda-Arroita2021}, \citealt{Casassus2021}). The correlation of AME amplitude and peak frequency with the dust temperature (T$_{\rm d}$) is tentatively reported in \cite{Poojon.2024}. Because the AME is directly related to the abundance of nanodust \citep{Hoang2016}, the potential explanation for the AME--T$_{\rm d}$ correlation is the increase in the abundance of VSGs due to rotational disruption of large grains by radiative torques in stronger radiation fields or higher T$_{\rm d}$ \citep{Hoangetal.2019}. Future observations and modelling taking into account the spatial variation of dust size distribution are needed to understand AME.

\subsection{Is AME polarised?}
\label{sec:Polarisation}

Polarisation of AME is a crucial probe to constrain the emission mechanisms and carriers of AME. The polarisation of AME, either due to spinning dust or magnetic dust is determined by the alignment of grains with the magnetic field.
\subsubsection{Spinning dust emission}

Spinning dust emission can be polarised if nanoparticles are aligned with the magnetic field. One important alignment pathway is resonance paramagnetic relaxation \citep{LazarianDraine2000}, which can produce polarisation levels of a few percent in the 20--30\,GHz range. In this mechanism, the polarisation direction is perpendicular to the magnetic field. Its efficiency depends on the spin-lattice relaxation rate, which for very small grains may proceed through Raman scattering of phonons; thus, the predicted polarisation is sensitive to grain composition and internal structure. Laboratory measurements of the relevant relaxation processes would be valuable for improving these estimates.

An independent process of spinning dust alignment is possible via a mechanism related to the dust interaction with radiation flow or gaseous flow that minimises the grain cross-section to the flow  \citep{Lazarian1995}. In \citet{HoangLaz.2018} the interaction of PAH particles and anisotropic radiation was considered. Particle were aligned in the positions with the minimal absorption cross-section. In the presence of magnetic field that induces fast precession of spinning dust magnetic moment about magnetic field, this would entail polarisation of the spinning dust radiation that depends on the angle between the magnetic field and the lines of sight.  The estimates of degree of polarisation range from 5 to 20 percent for the reflection nebulae dust to 3 percent for photo-dissociation regions. For spinning dust in cold neutral medium, the expected polarisation degree is less than 1 percent.

The overall polarisation of  spinning dust emission further depends on the intrinsic alignment efficiency of the nanoparticles and on magnetic-field turbulence along the line of sight. Nanoparticles can be aligned by paramagnetic relaxation \citep{davis:51}, and the exact degree of alignment depends on the grain rotation rate and magnetic susceptibility \citep{HoangLazMartin.2014}.
Nanosilicates with a large dipole moment have a slower rotation rate due to stronger electric dipole damping. As a result, the polarisation degree of spinning nanosilcates can be $\sim 20\%$ for the small dipole of $\beta=0.1D$, but it decreases to $\le 1\%$ for $\beta=1D$ \citep{Hoang2016}. In contrast, the diamagnetic nature of PAHs leads to much weaker alignment and correspondingly low polarisation levels. For the diffuse ISM, spinning PAH emission is expected to have polarisation fractions $\lesssim 3\%$ \citep{HoangLaz.2018}. Magnetic-field tangling will further suppress the observed polarisation. Based on the weak polarisation in the 2175 $\AA$ feature along two sight-lines towards two stars, HD 197770 and HD 147933-4, \cite{Hoang2013} derive the weak alignment of PAHs and constrain the polarisation fraction of $\leq 1\%$ for $\nu> 20$ GHz.

\subsubsection{Magnetic Dipole Emission}

The polarisation of magnetic dipole emission from iron nanoparticles depends strongly on whether they are free-flying grains or are inclusions in larger grains. \cite{Draine&Lazarian1999} and \cite{DraineHensley2013} demonstrated that perfectly aligned free-flying iron nanoparticles could achieve polarisation fractions of $\sim 30\%$. \cite{Hoang&Lazarian2016} calculated the alignment efficiency of magnetic particles as a function of size, finding that large grains ($\ge 1$ gm) are poorly aligned, whereas smaller grains attained high degrees of alignment and could produce emission polarised at up to $10-30\%$ levels. 

In the case of magnetic inclusions within a larger non-magnetic grain, the polarised emission depends both on the alignment of the grain and on the relative importance of the magnetic dipole emission from the inclusions and the electric dipole emission from the matrix of atoms, which are polarised orthogonally with respect to each other \citep{DraineHensley2013}. For randomly oriented magnetic inclusions in a silicate matrix, \citep{DraineHensley2013} found a drop in the polarisation fraction beginning at $\sim 10^{3}$ GHz (300\,$\mu$m) and extending to lower frequencies as the magnetic dipole emission becomes comparable to the emission from the silicate material. At low frequencies ($\sim 10$ GHz), the polarisation can even undergo a reversal, provided the magnetic Fe fraction is large enough. 

Thus, in the frequency range where both magnetic dipole emission and electric dipole emission from the grain are important ($\sim$ 10--100\,GHz), the polarisation fraction of the emission is low ($\le 5\%$). \cite{HoangLaz.2016a} investigated the alignment efficiency of grains with magnetic inclusions due to radiative torques, finding that the enhanced magnetic susceptibility due to the inclusions enabled the grain to achieve nearly perfect alignment due to the joint effects of enhanced magnetic relaxation, radiative torques, and collisional excitation. Thus, the polarisation fraction of the emission from large grains with magnetic inclusions is likely limited only by the degree to which the magnetic dipole and electric dipole emission processes are self-cancelling.


\section{Current Observational Status}
\label{sec:2-observations}

AME is detected in the same frequency range targeted by CMB experiments. As a result, CMB missions, particularly those mapping large sky areas or the full sky, have provided much of the available data on AME. In this section we separate the observational status into two categories: large-scale surveys and targeted observations of individual clouds.


\subsection{Large-Scale Missions}
\label{sec:observations-large}
Observations covering large areas of the sky provide extensive datasets that can be correlated with other wide-area surveys. Such data are particularly valuable for identifying global trends in the emission and for enabling statistical analyses that compare regions with different physical properties across many independent lines of sight.

\subsubsection{Large-Scale Datasets}
On large angular scales, measurements of the AME spectrum make use of data from \textit{Planck} (28--857\,GHz) and \textit{WMAP} (23--93\,GHz). These surveys generally only probe the high-frequency end of the AME spectrum, and they do not independently constrain the synchrotron or free–free emission. To isolate the AME from these underlying components and recover its spectral shape, complementary observations below 20\,GHz are required. 
Indeed, the Tenerife experiment at 10/15\,GHz \citep{2002ApJ...567..363D, deOliveira-Costa2004}, 19\,GHz data \citep{2003MNRAS.345..897B},  and the COSMOSOMAS experiment at 11--16\,GHz \citep{Watson2005, 2006MNRAS.370...15F} were instrumental in providing evidence for a turnover around 20\,GHz, supporting the spinning-dust hypothesis. As joint analyses are limited by the lowest common angular resolution, the AME is typically extracted at $1^{\circ}$ FWHM on large scales.

Historically, legacy total intensity surveys such as the 408\,MHz full-sky map by \citet{Haslam1982}, the 820\,MHz Dwingeloo northern-sky survey by \citet{Berkhuijsen1972}, and the 1.42\,GHz full-sky survey by \citet{Reich2001} that also included polarisation laid the groundwork for anchoring synchrotron and free–free emission. However, these maps suffer from systematics and incomplete beam characterisation, leading to large calibration mismatches between compact and diffuse emission \citep{Irfan2014a, Wilensky2024}. Reprocessed versions and scale-correction factors offer some improvement but cannot fully eliminate these issues. The 2.326 GHz HartRAO survey \citep{Jonas1998}, covering the southern sky with a well-characterised beam, represented an improvement in calibration and serves as a reliable baseline for the combined level of synchrotron and free–free in intensity.

The limitations of these historical surveys motivated a new generation of single-dish, ground-based instruments, namely S-PASS, C-BASS, and QUIJOTE, which measure both in intensity and polarisation and form the backbone of modern AME analyses. These datasets offer improved spectral coverage and control of systematics, allowing a cleaner separation of the AME.

\textbf{S-PASS} \citep{spass_release} covers the southern sky ($\delta \lesssim -1^{\circ}$) at 2.303\,GHz with an angular resolution of 8.9' using the 64\,m Parkes telescope, measuring both total intensity and polarisation. In intensity, it provides reliable constraints on the combined synchrotron and free–free emission.

\textbf{C-BASS} \citep{project_paper} has surveyed the northern sky ($\delta > -15.6^{\circ}$) at 4.76\,GHz in total intensity and polarisation with 44' resolution using a 6.1\,m  Gregorian telescope at the Owens Valley Radio Observatory. Its frequency lies just below the AME peak, making it a crucial reference for constraining synchrotron and free–free components and thereby isolating the AME amplitude. C-BASS and S-PASS together deliver a full-sky baseline for the combined level of synchrotron and free-free emission, enabling accurate constraints on the AME amplitude.

\textbf{QUIJOTE} \citep{Rubino-Martin2012b} covers the northern sky ($\delta \gtrsim -30^{\circ}$) at 10--40\,GHz in intensity and polarisation using two 2.25\,m telescopes at Teide Observatory with a resolution of 54'--18'. Public QUIJOTE-MFI maps at 11.1, 12.9, 16.8, and 18.8\,GHz \citep{mfiwidesurvey} uniquely sample the low-frequency side of the AME spectrum, constraining the AME peak frequency and width when combined with S-PASS and C-BASS.

Another large-scale project, though distinct from the surveys discussed above, is \textbf{COMAP} \citep{Cleary_2022, Rennie_2022}. The COMAP Pathfinder observes the inner Galactic plane ($|b| < 2^{\circ}$) from the northern hemisphere using a 10.4\,m Cassegrain telescope at the Owens Valley Radio Observatory. Operating between 26 and 34\,GHz with a 4.5' FWHM, it achieves higher angular resolution than existing single-dish surveys while retaining sensitivity to the large-scale emission that is typically inaccessible to interferometers. Its frequency range lies near the AME peak, making it especially valuable for correlating AME morphology with dust tracers and for targeted observations of specific regions. Planned extensions to COMAP will add receivers operating at 12--20 GHz. Joint fitting of the 12--34\,GHz continuum will (i) locate the AME spectral maximum to $\lesssim$1\,GHz accuracy, and (ii) measure the width and asymmetry of the emission bump.  Spatially-resolved measurements with these instruments of the peak frequency and low-frequency fall-off will translate directly into constraints on the nanograin size and rotational velocity distributions.

\subsubsection{Findings from Large-Scale Studies}

Overall, AME has been estimated to account for about half of the total intensity at 30\,GHz in the Galactic plane \citep{Planck2016XXV, planck_sd:13, Planck2015_X, Mateo2023}, making it one of the dominant components of microwave foreground emission.

A number of studies have explored the distribution and spectral characteristics of AME. The first large-scale statistical analysis was carried out by \citet{planck_sd:13}, who combined \textit{Planck} and legacy low-frequency data to study 98 compact clouds, finding significant AME in 42 of them. This work established the strong correlation between AME and thermal dust emission and demonstrated that AME is not spatially uniform, but instead varies in emissivity and peak frequency across different environments. 

Subsequent studies incorporating QUIJOTE data between 10 and 20\,GHz improved the characterisation of the AME spectrum. \citet{Poidevin2023} identified 44 significant AME sources and provided tighter constraints on the peak frequency, while \citet{Mateo2023} extracted AME emission along the inner Galactic plane ($|b| < 10^{\circ}$) on a pixel-by-pixel basis by defining a consistent background reference across datasets. These works further emphasised that local environmental conditions strongly influence the emissivity and spectral properties of AME. However, without C-BASS and S-PASS data, constraints on the combined synchrotron and free-free emission remained limited, reducing the accuracy of the AME separation.

Building on these studies, \citet{CepedaArroita2026} recently combined QUIJOTE, C-BASS, and S-PASS data to derive fully constrained AME amplitudes, peak frequencies, and widths for 144 Galactic clouds, significantly expanding the sample size and improving constraints on peak frequency and width. The study found that AME spectra are generally broader than predicted by spinning dust models, suggesting either multiple spinning dust components along the line of sight or an incomplete representation of the grain size distribution in current models. Nonetheless, the narrowest observed widths align with theoretical expectations, offering new evidence in favour of the spinning dust hypothesis. The amplitude of AME was found to correlate most strongly with thermal dust peak flux or radiance over three orders of magnitude. A sub-linear scaling between AME and thermal dust radiances and amplitudes implies that AME becomes less efficient in denser environments, consistent with the depletion of small grains through adsorption onto larger ones. A tight positive correlation was also identified between the AME peak frequency and the temperature of large dust grains, a relationship not readily reproduced by current spinning dust models. The authors argue that this trend likely reflects how the local radiation field affects the grain size distribution, highlighting the need for models that capture the interdependence between environmental conditions and the evolution of grain properties.


\subsection{Targeted Galactic Observations}

Targeted observations offer higher angular resolution and greater sensitivity than wide-area surveys. Current interferometers, such as the JVLA and ALMA, can map AME with enough detail to enable direct comparison with mid-infrared templates that trace the population of very small grains. High-resolution observations of individual nearby clouds hosting PDRs are particularly valuable, as they can be compared with models of the physical conditions across the transition from the dense molecular material to the more diffuse gas closer to the ionising source. This allows the spatial distribution of AME to be linked to the changing physical conditions across the PDR gradient.

\subsubsection{Targeted Example: $\rho$\,Ophiuchi}

The $\rho$~Ophiuchi molecular cloud lies in the Gould Belt, at a distance of $134.3 \pm 1.2$\,pc \citep{Gaia2018}. It is one of the brightest AME sources in the sky \citep{Casassus2008,planck_sd:11}. The UV radiation from the star HD\,147889 gives rise to a filamentary PDR, whose brightest IR component is $\rho$\,Oph\,W \citep{Liseau1999,Habart2003}.

AME was first detected in this region by the Cosmic Background Imager (CBI; \citealt{Padin2002}), a 13-element interferometer located at the Chajnantor plateau in Chile (the current ALMA site). The CBI had 10 spectral channels covering 26--36\,GHz. Its data revealed abundant AME at 31\,GHz, originating from $\rho$\,Oph\,W \citep{Casassus2008,Arce-Tord2020}. Later, observations from the {\em Planck} spacecraft \citep{planck_sd:11} showed that $\rho$\,Oph has the highest AME emissivity per column density among all AME sources detected by {\em Planck}. Moreover, on angular scales of $\sim$1$^{\circ}$, it shows no significant free-free or synchrotron emission compared to the AME, making it one of the few known regions where AME appears to dominate the microwave continuum emission.

\begin{figure}[!t]
 \centering
 \includegraphics[width=\textwidth]{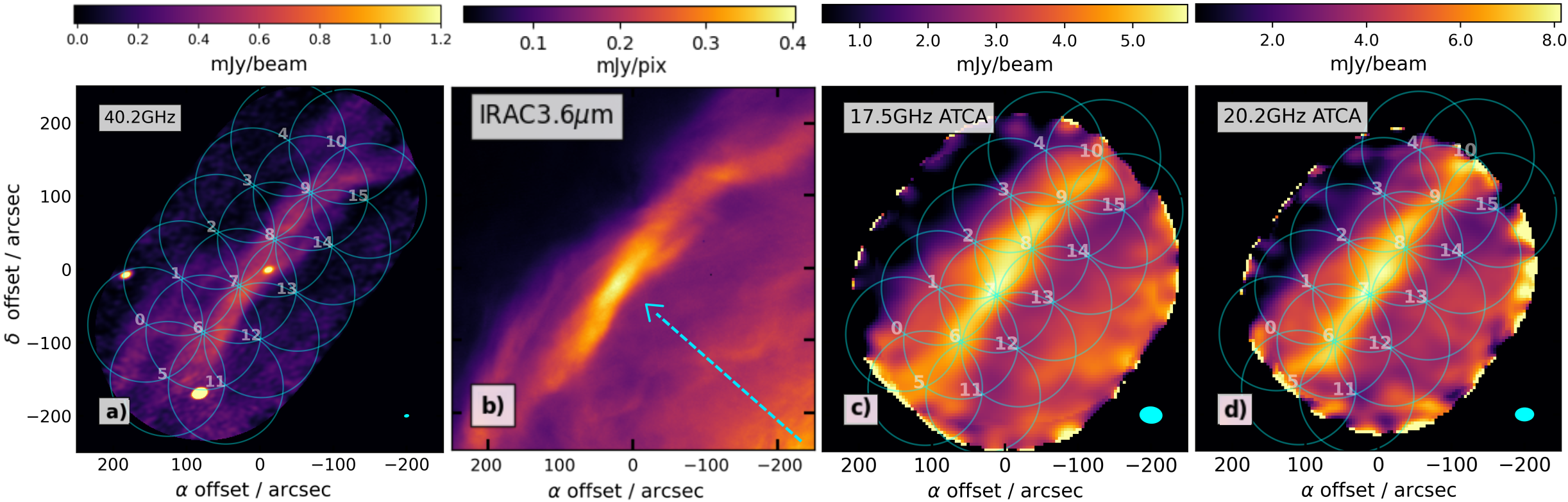}
 \caption{Comparison between the ALMA Band~1 mosaic at 40.2\,GHz (a), the \textit{Spitzer} mid-IR map at 3.6\,\textmu m (b), and the Australia Telescope Compact Array (ATCA) maps at 17.5\,GHz (c) and 20.2\,GHz (d), overlaid with the ALMA Band~1 pointings. Note the striking correlation between the 40.2\,GHz emission and the IRAC 3.6\,\textmu m map. The emission observed in the ATCA maps (c and d) appears shifted toward the north-east compared with the ALMA map, effectively tracing regions further inside the PDR. Figure adapted from \citet{Casassus2026}. }
 \label{fig:roph}
\end{figure}

The region around the peak emission in the $\rho$\,Oph\,W PDR was studied at higher angular resolution ($\sim$7$''$) using ALMA at 40\,GHz. Figure~\ref{fig:roph} shows, on the left, the ALMA mosaic of $\rho$\,Oph\,W at $\sim$7$''$ resolution, alongside the {\em Spitzer} 3.6\,\textmu m map and the ATCA maps at 17.5\,GHz and 20.2\,GHz. There is a strong correlation between the ALMA and 3.6\,\textmu m emission. Interestingly, the peak of the ALMA emission does not coincide with the brightest regions in the ATCA maps, where the emission is shifted toward the north-east, i.e. deeper inside the molecular cloud.

This shift can be interpreted within the spinning dust framework as a change in the grain-size distribution: smaller grains, spinning faster, dominate at higher frequencies in the outer parts of the PDR facing the ionising stars, while larger grains emit at lower frequencies deeper into the cloud. However, confirming this interpretation requires verifying that the 17--40\,GHz emission is indeed AME, free from low-frequency contamination at arcsecond angular scales. A possible contaminant is free-free emission arising from C$^+$ recombination with free electrons (C\,\textsc{i} continuum) in the cold plasma of $\rho$\,Oph\,W. A low-frequency map with comparable angular resolution would be instrumental in distinguishing these components unambiguously.


These ALMA observations, with arcsecond resolution, demonstrate that such angular resolution is essential to resolve the morphological variations required to understand the nature of AME. To fully exploit these data, it is crucial to characterise the low-frequency emission components that anchor the AME spectrum. Observations with SKA-Mid in the $\sim$1--15\,GHz range, with comparable (arcsecond) angular resolution and high surface-brightness sensitivity, will enable this by mapping the same region over a large field of view. The combination of SKA-Mid and ALMA data will allow a direct, spatially resolved comparison between low-frequency emission, high-frequency dust-related emission, and mid-infrared morphology.


\subsection{Extragalactic Detections}
\label{sec:extragalactic}

Despite the wide range of Galactic environments where AME is observed, extragalactic detections remain scarce. Early searches using \textit{WMAP} and \textit{Planck} data were largely inconclusive, although a sub-mm excess in the Magellanic Clouds was interpreted as AME \citep{Israel2010,Bot2010}, and a first likely detection was reported in a star-forming region of NGC\,6946 \citep{Murphy2010,Scaife2010}. Follow-up studies found AME emissivities significantly higher than in the Milky Way, challenging standard spinning-dust models \citep{Hensley2015}.

\begin{figure}[!t]
 \centering
 \includegraphics[width=0.5\textwidth,angle=0]{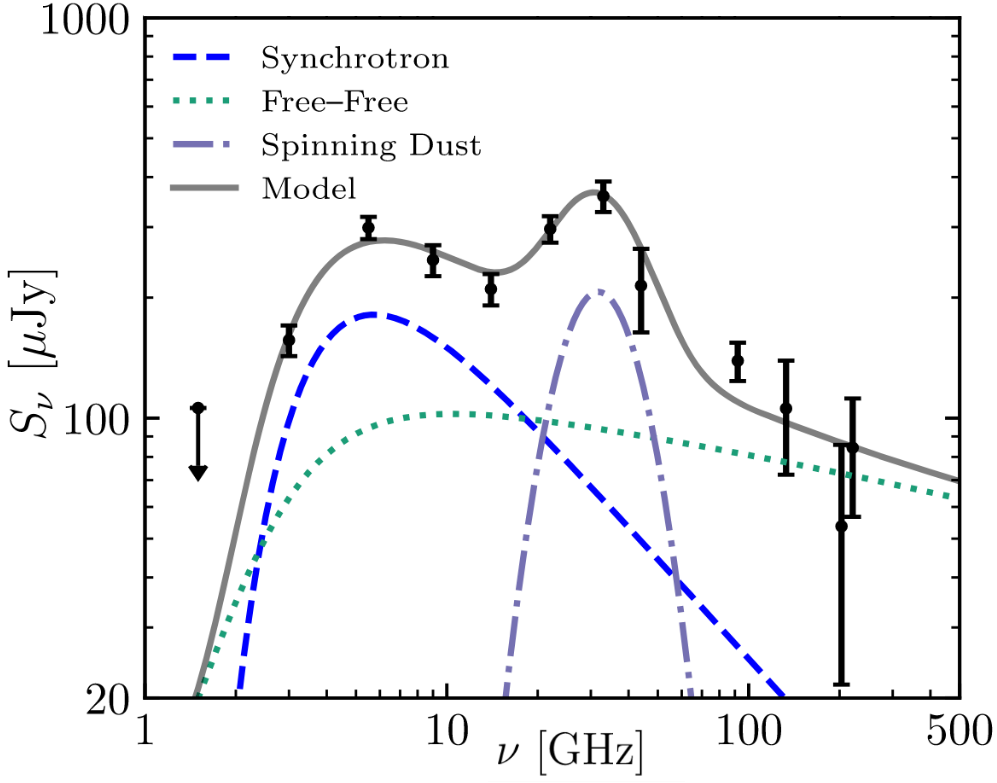}
 \caption{Radio-to-millimeter spectrum of NGC\,4725\,B, displaying a strong peak at $\approx$30\,GHz. Figure adapted from \citet{Murphy2020}.}
 \label{fig:NGC4725B}
\end{figure}

A more extreme case is NGC\,4725\,B, which exhibits a strong $\sim$30\,GHz peak with little associated sub-mm dust emission \citep{Murphy2018,Murphy2020} (Fig.~\ref{fig:NGC4725B}). The implied AME-to-dust ratio is orders of magnitude higher than in the Galaxy, difficult to reconcile with spinning-dust models without invoking extreme grain properties, and not fully explained by magnetic dust scenarios. Higher-resolution observations are required to clarify the origin of this emission.

Recent surveys have expanded the sample: \citet{Linden2020} reported 33 AME candidates in nearby galaxies ($d_L < 30$\,Mpc) using JVLA observations between 3--33\,GHz, while \citet{Dage2025} identified compact radio counterparts in several cases, suggesting an origin in young, compact \hii\ regions.

A notable difference with Galactic AME is that extragalactic detections are predominantly associated with compact regions, whereas Galactic AME is typically diffuse. This may reflect observational biases or the difficulty of isolating AME from other emission mechanisms in external galaxies.

SKA-Mid will significantly advance this field by providing the sensitivity and angular resolution required to detect faint AME components, separate them from synchrotron and free-free emission, and resolve their spatial association with dust. This will enable robust tests of AME models across galactic environments.

\subsection{Polarisation Upper Limits}
\label{sec:polame}

Establishing whether AME exhibits any level of polarisation is crucial for experiments aiming to detect the elusive B-mode signal (the {\em smoking-gun} evidence of cosmic inflation) \citep{Ade2019,Abazajian2022,LiteBIRD2023}. As shown by \citet{remazeilles2016}, neglecting an AME component with a polarisation fraction as small as $\sim$1\% could introduce a significant bias in the estimation of the {\em tensor-to-scalar ratio} ($r$), a key parameter that quantifies the amplitude of gravitational-wave fluctuations generated during inflation.

Regarding theoretical models for AME, it has been proposed that quantum-mechanical effects may suppress grain alignment, leading to a very low level of polarisation if the AME arises from spinning dust grains \citep{DraineHensley2016}. On the other hand, most models of magnetic dipole emission \citep{draine:99,DraineHensley2013,Hoang&Lazarian2016,Hoang2016} predict high polarisation fractions ($\gtrsim 10\%$). Therefore, measuring the polarisation level of AME could help to discriminate between different emission mechanisms.

There have been numerous attempts to measure the polarisation fraction ($\Pi = P/I$) of AME, all on angular scales of order $\sim$1\degree.  The most stringent upper limits on $\Pi$ for AME are $\sim$0.5\% \citep{Dickinson2011,lopez-caraballo:11,rubino-martin:12, 2017MNRAS.464.4107G}, indicating that the polarisation level of AME is low at large angular scales.

Since the 2018 review by \cite{Dickinson2018}, the 
most stringent polarisation limits have been reported by \cite{gonzalezgonzalez2025} for the  
$\rho$-Ophiuchi, Perseus, and W43 regions. These limits are $\Pi_{\rm AME} < 1.1$\,\% at 28.4\,GHz for $\rho$-Ophiuchi;  $\Pi_{\rm AME} < 1.1$\,\% at 22.8\,GHz for Perseus; and $\Pi_{\rm AME} < 0.28$\,\% at 33\,GHz for W43. Those constraints are already 
approaching the level at which intensity-to-polarisation leakage must be controlled below 0.1 per cent in order to achieve further improvement.
Regarding constraints on diffuse AME emission, \cite{herman25} used \textit{Planck} LFI and \textit{WMAP} polarisation 
data to set upper limits at the percent level. In particular, assuming a synchrotron spectral index prior of $\beta_{\rm S} < -3.1$, they find $p_{\rm AME}^{\rm max} \lesssim 0.6$\,\% (95\,\% confidence). 

\begin{figure}[!t]
 \centering
 \includegraphics[width=\textwidth,angle=0]{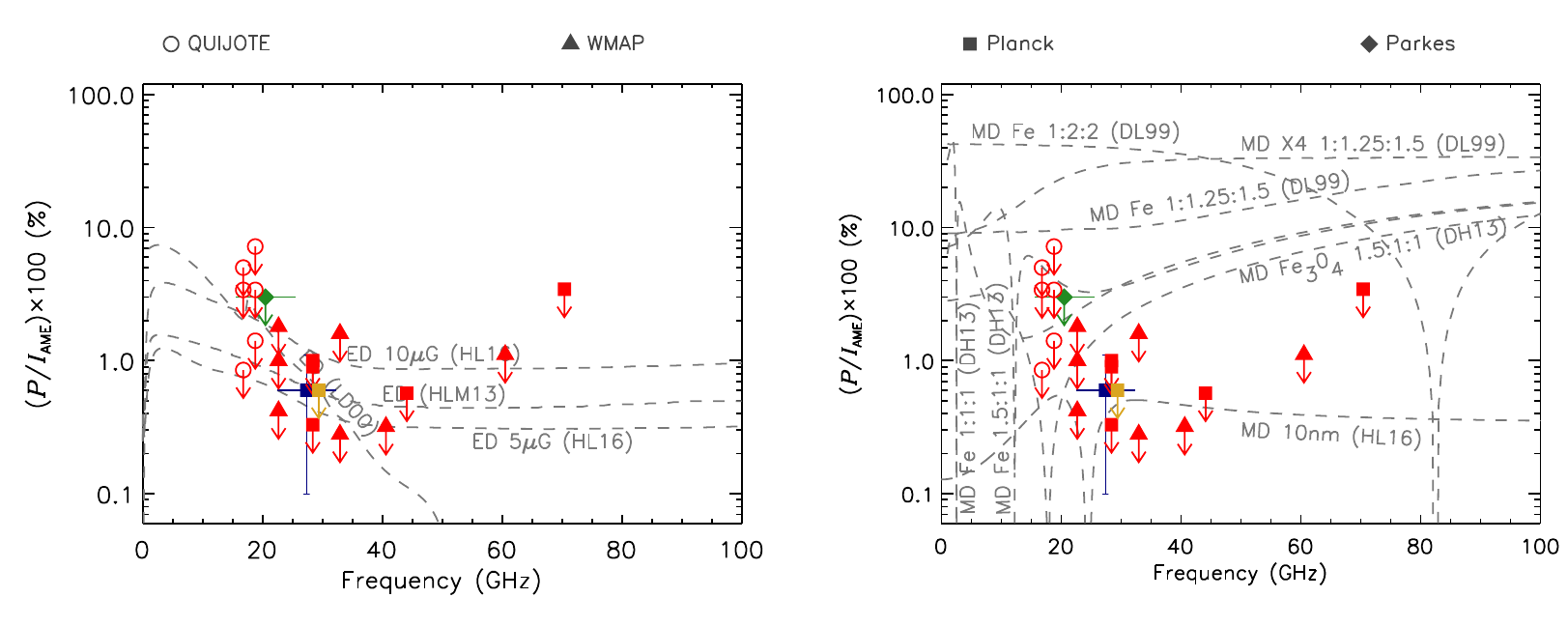}
\caption{Recent constraints on the AME polarisation fraction from different experiments (see legend), compared with model predictions (grey dashed lines) for electric dipole (ED; left) and magnetic dipole (MD; right) emission. Measurements from \cite{Battistelli2015} and \cite{gonzalezgonzalez2025} (compact regions) are shown in green and red, while large-scale constraints from \cite{Planck2016XXV} and \cite{herman25} are shown in blue and yellow. Some polarisation models for MD are ruled out by the measurements. }
\label{fig:pol_constraints}
\end{figure}

Figure~\ref{fig:pol_constraints} presents a summary of current upper
limits on AME polarisation, updated from \citet{Dickinson2018}.

\section{The Role of the SKAO, Expected Outcomes \& Impact}
\label{sec:3-SKA}

The SKAO will play a pivotal role in advancing AME science by probing the $\sim 1$--$15$\,GHz range, which remains poorly explored and has so far been observed mainly at $\sim1^{\circ}$ resolution by CMB experiments (see Sec.~\ref{sec:observations-large}). Its combination of sub-arcsecond resolution and high sensitivity will enable measurements of the AME SED on the rising part of the spectrum, breaking degeneracies in spinning-dust models that are currently constrained only above the peak.

In the following, we present simulations and expected outcomes for AME studies with the SKAO, including Galactic emission, protoplanetary disks, extragalactic sources, polarisation, and emission modelling, as well as synergies with other facilities.

We focus on continuum measurements, which provide the most direct probe of AME, as spinning dust produces a quasi-continuum spectrum rather than discrete lines. Previous searches for rotational line emission from quasi-symmetric PAHs have yielded no detections \citep{Ali-Haimoud2015,Casassus2026}. The spectral capability of SKA-Mid nonetheless allows simultaneous searches for molecular lines, including those from complex organic molecules or substituted PAHs in dense environments, providing complementary constraints on gas-phase conditions.

A key challenge is component separation and control of observational systematics, particularly in crowded regions and for polarisation. Multiple emission mechanisms overlap spectrally and spatially, while interferometric limitations (e.g. incomplete $uv$-coverage, missing large-scale emission, and calibration uncertainties) can bias SEDs. SKA-Mid mitigates these effects through its wide bandwidth, high sensitivity, and angular resolution, enabling robust spectral decomposition and reduced source confusion, while polarisation systematics can be addressed through careful calibration and high-resolution observations.

We focus on SKA-Mid (0.35--15.4\,GHz), which probes the AME spectrum near its peak, whereas SKA-Low (50--350\,MHz) lies well below this regime and is dominated by synchrotron emission. Unless otherwise stated, results assume the AA$^*$ configuration, which already provides sub-arcsecond resolution and sub-$\mu$Jy\,beam$^{-1}$ sensitivity at GHz frequencies, sufficient for the science goals described here.

\subsection{Galactic Emission}

Nearby Galactic objects are ideal testbeds for investigating the emission mechanisms responsible for AME. Angular resolutions of $\sim1$--$3$\,arcsec are required to resolve the structure of AME in nearby PDRs and enable direct comparison with mid-IR and sub-mm tracers of small grains; these are readily achieved by SKA-Mid, which delivers sub-arcsecond resolution at GHz frequencies. Its wide frequency coverage across $\sim1$--$15$\,GHz further allows competing emission mechanisms (synchrotron, free--free, and spinning dust) to be disentangled through their distinct spectral behaviour. It will also allow studies of AME in compact environments, including protoplanetary disks.

\subsubsection{Diffuse emission: : $\rho$\,Ophiuchi}

Diffuse emission from the ISM follows power-law spatial power spectra, with most of the signal distributed over large angular scales. The dense $uv$-coverage and high surface-brightness sensitivity of SKA-Mid, however, enable robust recovery of this emission when appropriate weighting and $uv$-tapering are applied, while still reaching arcsecond-scale resolution. Although the AME spectrum in PDRs decreases toward lower frequencies, the sensitivity of SKA-Mid allows the rising part of the spectrum to be probed with high fidelity.

We assess the observability of excess microwave emission (EME) in the $\rho$\,Oph\,W filament, a well-characterised PDR with extensive ancillary data. The reported linear correlation between IRAC\,3.6\,$\mu$m and ALMA Band\,1 \citep{Casassus2026}, together with ATCA observations at 17\,GHz \citep{Casassus2021}, provides an empirical estimate of the 3.6\,$\mu$m-correlated signal, which we extrapolate to SKA-Mid\,5b assuming a $\nu^2$ spectrum.

\begin{figure}[!t]
 \centering
 \includegraphics[width=1\textwidth]{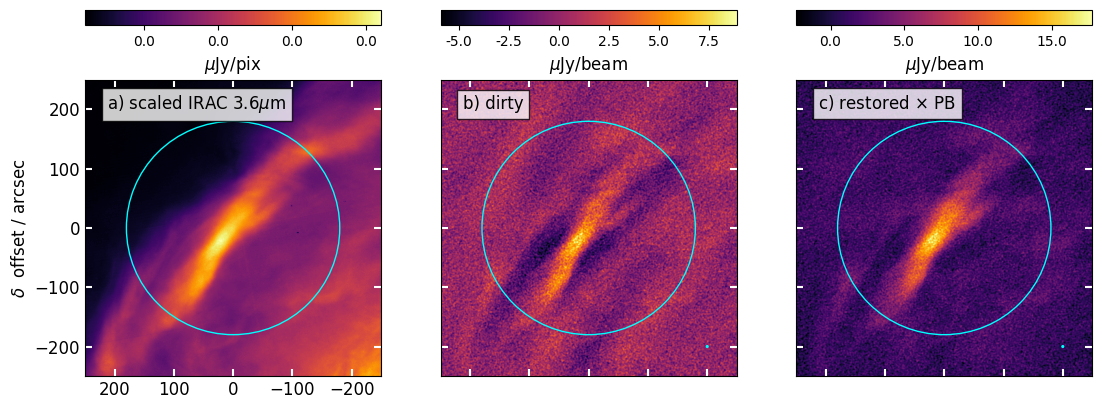}
 \caption{Simulation, for a single pointing,  of the expected signal from the $\rho$\,Oph\,W
   filament at 11.85\,GHz (SKA-Mid\,5b) and in AA$^*$
   configuration. $x-$ and $y-$ axis correspond to offset R.A. and
   Dec., in arcsecs. The primary beam is shown in all images as a single cyan contour at half-maximum.
   The input IRAC\,3.6\,$\mu$m sky image, in a), has
   been scaled from the observed linear slope with ALMA\,Band\,1, and
   then extrapolated to ATCA\,17\,GHz and then to 11.85\,GHz. The
   dirty map, in b), corresponds to a 3\,h integration with Briggs
   weighting $r=0.5$ and a 7\,arcsec taper, giving a fairly circular
   $3.0$\,arcsec Gaussian beam. The noise level in b) is
   0.7\,$\mu$Jy\,beam$^{-1}$. The restored image, in c), was obtain
   with the {\tt skymem} package \citep[][]{Casassus2021} in a
   pure-$\chi^2$ fit, and without image priors.}
 \label{fig:sims-roph}
\end{figure}

Simulated SKA-Mid visibilities were generated with {\tt pyralysis} and {\tt ska-ost-array-config}, and imaged using non-parametric methods ({\tt gpu-uvmem}, {\tt skymem}). Adopting a $uv$-taper to obtain a $3$\,arcsec beam and a system temperature of 14.5\,K, we recover the expected thermal noise level of $\sim70$\,nJy\,beam$^{-1}$ in a 3\,h integration. The reconstructed images recover most of the input signal from the $\rho$\,Oph\,W filament (Fig.~\ref{fig:sims-roph}). These observations demonstrate that SKA-Mid in AA$^*$ can map diffuse AME in PDRs at arcsecond resolution, enabling direct measurements of its spatial distribution and correlation with small-grain tracers, and thereby providing a critical test of spinning-dust models.

Recent joint radio-infrared modelling of the $\rho$\,Oph\,W PDR, combining {\em Spitzer} and {\em Herschel} data with ALMA and ATCA continuum measurements, shows that data below $\sim20$\,GHz are essential to constrain spinning-dust parameters and break degeneracies in nano-grain models. The infrared emission traces small-grain properties, while the rising radio SED anchors the spinning-dust component and constrains key parameters such as grain size distribution, electric dipole moments, and local conditions. This underscores the importance of SKA-Mid for AME studies \citep{Casassus2026b}.

\subsubsection{Protoplanetary Disks}
\label{sec:compact}

SKA-Mid observations could reveal the  largest dust population in protoplanetary disks and constrain grain and pebble growth. But, because of the steeply falling Rayleigh-Jeans spectrum, such  observations of dust are difficult. Additional continuum components become important below $\sim 40$\,GHz. Excess microwave emission (EME) over the Rayleigh-Jeans dust extrapolated from mm-wavelengths, is routinely observed in unresolved data, both in HAeBe and T\,Tauri stars \citep[][]{Skinner1993ApJS...87..217S, Ubach2012MNRAS.425.3137U, Ubach2017MNRAS.466.4083U}. Such EME cannot entirely be due to a population of very large grains, with negligible emissivity at mm-wavelengths and shorter, as the required dust masses would correspond to disk masses comparable to that of the star \citep[][]{Natta2004A&A...416..179N,2005ApJ...626L.109W}.

The stellar H\,{\sc i} free-free emission also kicks-in at cm-wavelengths or shorter \citep[][]{Skinner1993ApJS...87..217S, Ubach2012MNRAS.425.3137U}. In fact, most sources turn out to be variable at cm-wavelengths, with flux density variations ranging from 30\% up to a factor of three \citep[][]{Ubach2017MNRAS.466.4083U}, thwarting measurements of the large dust mass reservoir in unresolved data, without an independent means to subtract the variable free-free component. The jets at the roots of the large-scale stellar outflows also have free-free counterparts \citep[][]{Carrasco-Gonzalez2019ApJ...883...71C}. Another possible, though not yet firmly identified extended free-free component may be due to the excess UV radiation from photospheric radiation, which generates a Str\"omgren surface akin to a thin H\,{\sc ii} region, thought to be the launching region of thermally driven photo-evaporative disk winds. Such H\,{\sc ii} regions at the disk surfaces may contribute an extended signal at long wavelengths.

In addition to the Rayleigh-Jeans dust and free-free components, spinning-dust emission could also be relevant in disks.  Spinning dust models tailored to T\,Tauri and HAeBe disks predict that this mechanism may significantly contribute to disk SEDs \citep[][see right panel of Fig. \ref{fig:SD-emm}]{Rafikov2006ApJ...646..288R, Hoang2018}. Even a small spinning dust contribution may mimic the unresolved SED expected from large grains \citep[][]{Hoang2018}. There are already several unresolved EME detections in disks, with SEDs matched by simple spinning dust models \citep[][]{Greaves2018NatAs...2..662G,   Greaves2022MNRAS.513.3180G, Chung2025}, including a prominent signal in PDS\,70 \citep[][]{Liu2024ApJ...972..163L}.

In order to separate each radio continuum component it is necessary to resolve the disks, and at the very least separate the stellar component from the disk signal, requiring angular resolutions finer than $\sim$100\,mas, and within reach of AA*. Coverage across $\sim1$--$15$\,GHz is essential to disentangle free--free, spinning dust, and Rayleigh--Jeans dust through their distinct SEDs and spatial distributions. Due to the dust-trapping mechanism, progressively larger grains that have not drifted onto the star are expected to pile-up in local pressure maxima, such as narrow rings or crescent-shaped anticyclonic vortices \citep[e.g.][]{Dullemond2018ApJ...869L..46D,Casassus2019MNRAS.483.3278C}. Rayleigh-Jeans dust is thus expected to be brighter in such localised dust traps. In contrast, spinning dust stems from very small grains that are well coupled with the gas. Therefore Rayleigh-Jeans dust should be compact and structured, while spinning dust should be diffuse.


Spinning dust  in disks  complicates the interpretation of radio data, especially close to cm-wavelengths. But it also provide a new window on the very small grain population. Photoelectric heating from PAHs regulates the temperature at the disk surface, and hence the rate of photo-evaporation \citep[][]{Ercolano2022EPJP..137.1357E}. However, PAHs abundances in disks are notoriously difficult to estimate from the IR aromatic bands, even when using JWST observations \citep[][]{Bandyopadhyay2025}. Radio observations with SKA-Mid could thus yield the first solid estimates of PAHs abundances in disks.



\subsection{Extragalactic AME}

SKA-Mid will also enable sensitive searches for AME in external galaxies. These measurements up to 15.4\,GHz will probe frequencies close to the typical AME peak, which on average lies at $21.9\pm 3.7$\,GHz \citep{CepedaArroita2026}. A detailed discussion of the SKAO's role in this topic can be found in the accompanying chapter by \citet{Yoon01.2026.SKA}.

A key requirement for extragalactic AME studies is the ability to resolve individual star-forming regions, where the emission is thought to originate. For nearby galaxies ($d \sim 10$--30\,Mpc), this corresponds to physical scales of $\sim$50--150\,pc, requiring angular resolutions of $\sim$0.3--1\,arcsec. Such resolution is achievable with SKA-Mid, in AA* configuration, across much of its frequency range, particularly above a few GHz, enabling the separation of compact AME-emitting regions from surrounding synchrotron and diffuse free-free emission.

In addition to its angular resolution, SKA-Mid will provide a substantial sensitivity improvement over current facilities such as the JVLA. For typical continuum observations at GHz frequencies, SKA-Mid is expected to reach sub-$\mu$Jy\,beam$^{-1}$ sensitivities (e.g. $\sim$0.8\,$\mu$Jy\,beam$^{-1}$ in $\sim$2\,hr integration), whereas the JVLA typically achieves $\sim$3.5\,$\mu$Jy\,beam$^{-1}$ under comparable conditions. This improvement will enable the detection of significantly fainter AME components in individual star-forming regions. This is particularly important in external galaxies, where AME is expected to be sub-dominant and easily obscured by stronger radio emission processes. The combination of high sensitivity, dense $uv$ coverage, and wide frequency bandwidth will therefore enable robust spectral decomposition and spatially resolved studies that are not currently feasible.

More distant galaxies may also see the AME peak redshifted into SKA-Mid bands, where it could contribute to their integrated spectral energy distributions. This may be particularly important when using radio continuum measurements to isolate free-free emission and derive star formation rates, as even a modest AME contribution could bias such estimates if not properly accounted for.


\subsection{Parameter Estimation from Spinning Dust Emission}

The new generation of spinning-dust models, from the original
\texttt{spdust} to the generalised \texttt{SpyDust} framework,
provides a physically grounded basis for interpreting AME spectra.

SKA-Mid Bands~5a (4.6--8.5\,GHz) and 5b (8.3--15.4\,GHz) will provide
continuous coverage of the low-frequency side of the spinning-dust
spectrum with $\sigma_I \approx 1.2$\,$\mu$Jy\,beam$^{-1}$ noise at
arcsecond resolution in a 1-hour integration \citep{Braun2019}.  This
is a major advance over the most comprehensive AME spectral study in
this frequency range to date \citep{CepedaArroita2026}, which rely on
C-BASS ($\Delta\nu = 0.49$\,GHz at 4.76\,GHz) and four QUIJOTE\,MFI
channels ($\Delta\nu = 2.2$\,GHz each at 11--19\,GHz), all at
$\sim$1\degree\ resolution with $\sim$5\,\% calibration
uncertainties. Combined with ALMA Band~1 (35--50\,GHz) and the ngVLA
\citep[1.2--116\,GHz;][]{Murphy2018-ngVLA}, SKA-Mid will enable
densely sampled SEDs from a few GHz through the spectral peak,
mitigating degeneracies between grain size distribution, dipole
moment, and environmental parameters that persist with current sparse
frequency sampling.

Although the mapping from observed SEDs
to ISM environment parameters remains highly degenerate, emerging
statistical techniques, such as moment-expansion approaches and
  global sensitivity analysis \citep{Zhang25_Spydust,
  Zhang26emsemble}, can help extract key parameter responses and
exclude unphysical parameter combinations. Together, these
multi-frequency observations and advanced inference tools will turn
spinning dust models into predictive probes of the interstellar
medium.

\subsection{Constraining Polarisation}

As described in Sect.~\ref{sec:polame}, all  current AME polarisation constraints have been obtained at angular scales of order $\sim$1\degree. However, these measurements cannot be extrapolated to smaller angular scales. The mixing of different polarisation vectors along the line of sight produces a depolarisation effect. Another difficulty in measuring the polarisation fraction at coarse resolution is that different total intensity emission mechanisms can be blended along the same line of sight. Since AME typically originates at PDRs, they are normally located near \hii\ regions, so free-free emission often contributes to the total intensity measured within a 1\degree\ beam encompassing AME regions.

Polarisation observations of AME with arcsecond-scale resolution with SKA-Mid up to 15.4\,GHz could alleviate these issues, as they can be targeted towards {\em pure} AME sources, reducing contamination in total intensity and mitigating beam depolarisation. 
Moreover, observations provided by SKA-Mid also open the possibility of exploring polarisation limits in extragalactic objects.

For a dual-polarisation system, the Stokes~$Q$ and $U$ sensitivities 
are comparable to the total intensity sensitivity.
Using the Band~5b sensitivity from \cite{Braun2019}, a deep 100-hour 
observation yields $\sigma_Q \approx 0.12\,\mu$Jy\,beam$^{-1}$.
Discriminating spinning dust from magnetic dipole emission 
(MDE; $\Pi \sim 10$--$30$\,\% for free-flying magnetic nanoparticles; 
\citealt{DraineHensley2013,Hoang&Lazarian2016})
requires constraining $\Pi$ to the percent level or below.
For spinning PAH emission, 
$\Pi \lesssim 3$\,\% is predicted in PDRs, decreasing to $\lesssim 1$\,\% 
in the diffuse ISM \citep{Hoang2013,HoangLaz.2018}. 
To achieve a 3$\sigma$ upper limit of $\Pi < 1$\,\%, 
a source must have a total intensity 
$I \gtrsim 3\sigma_Q / 0.01 \approx 36\,\mu$Jy\,beam$^{-1}$ 
in a 100-hour Band~5b observation. 
For brighter sources with $I \gtrsim 120\,\mu$Jy\,beam$^{-1}$, 
the constraint tightens to $\Pi < 0.3$\,\%, matching the most stringent 
degree-scale upper limits currently available 
\citep[$\Pi_{\rm AME} < 0.28$\,\% for W43;][]{gonzalezgonzalez2025}, 
but now at arcsecond resolution where beam depolarisation is negligible 
and contamination from adjacent free-free emission can be spatially separated.


\subsection{Synergies with Other Facilities}\
SKA-mid will eventually cover the 350--15400\,MHz frequency range, directly probing the spectral regime where anomalous microwave emission (AME) peaks and where its separation from synchrotron and free-free emission is most critical. As such, SKA-mid will provide the primary observational constraints on the AME spectrum and morphology, forming the foundation for interpretation with complementary facilities.

At the array's highest resolution configuration (0.08$''$ at Band 5-b, AA*),it will provide the optimal counterpart to ALMA observations of AME, particularly useful in compact objects such as protoplanetary disks and extragalactic targets. Additionally, as the baseline lengths within the core array are around 1 km, this allows for imaging at coarser resolutions: between 0.08 and 3.6\,arcmin. This multi-scale capability is particularly important when connecting compact AME sources to their surrounding environments.

The next generation of CMB experiments will observe the sky at resolutions of $\sim1$ arcmin \citep{so, cmbs4} focusing on the 30--300\, GHz frequency range. The combination of the SKA-mid and next-generation CMB observations would enable investigations of arcmin AME structures. Specifically, partial sky maps at radio and microwave frequencies can be jointly used to constrain foreground (free-free, synchrotron, AME and thermal dust) spectral and spatial behaviour to high precision. In \citet{irfan15} the authors introduce 5\,GHz C-BASS data to help constrain free-free and synchrotron emission within multi-experiment spectral energy distributions across several compact Galactic plane regions and identify a new source of compact AME: W44. The SKA-mid will provide direct measurements of AME at frequencies $> 10$\,GHz, while simultaneously constraining free-free and synchrotron emission across the radio regime, enabling the spectral coverage required for robust component separation.


When used as an interferometric array the SKA-mid is limited to probing the sky at resolutions smaller than several arcmins; any angular scales larger than the minimum baselines can measure will be absent from the images. However, the SKAO precursor, MeerKAT, has demonstrated that its dishes can be used in single-dish mode, as opposed to as part of an interferometric array. In \citet{mk22} the cosmology experiment MeerKLASS used single-dish L-band MeerKAT data to investigate diffuse Galactic emission under 1075\,MHz.

The 13.5\,m diameter MeerKAT dishes will be combined alongside 15m diameter dishes to form the full SKA-mid array.  SKA-mid band 5a (4.6 $-$ 8.5\,GHz) and 5b (8.3 $-$ 15.4\,GHz) will have resolutions of 9 $-$ 17\,arcmin and  5 $-$ 10\,arcmin, respectively. When used in single-dish mode, all angular scales larger than the beam FWHM are preserved. Therefore single dish 5a and 5b observations can be smoothed to degree scales and used alongside QUIJOTE, \emph{WMAP}, \emph{Planck} and next-generation CMB observations \citep{basu19, skacmb} to understand the large scale distribution of AME across our Galaxy with respect to both diffuse and compact sources of Galactic emission. A prototype 15m SKA-MPIfR (SKAMPI) dish for SKA-mid was assembled on-site at the end of 2018, achieved first light in 2023. SKAMPI is currently undergoing commissioning whilst mapping the southern hemisphere at S-band in single-dish mode \footnote{\url{https://www.skao.int/en/news/512/ska-mid-prototype-dish-creates-first-light-image} with observational access given to the South African and German science communities. Given the successes of SKAMPI and MeerKLASS, and the need for 5 - 15\,GHz constraints on the low frequency end of the AME spectral form, we would strongly support an extension of SKAO functionality to include the single-dish data reduction software and observational strategies established by these pioneering surveys.}      

\paragraph{SPHEREx} 

The {\em SpectroPhotometer for the History of the Universe, Epoch of Reionization, and Ices Explorer} (SPHEREx) is a NASA mission designed to perform the first all-sky near-infrared spectroscopic survey. During its nominal 2-year mission, SPHEREx will construct an all-sky spectral map of stars, galaxies, and diffuse emission, obtaining 0.75--5\,$\mu$m low-resolution ($\lambda/\Delta\lambda \sim 35\text{--}130$) spectra of every $6.2\times6.2$\,arcsec$^2$ pixel on the sky \citep{Bock2026}.

SPHEREx will provide blended measurements of the 3.3 and 3.4\,$\mu$m emission bands, prominent features associated with PAHs. These measurements provide key information on the evolution of the aliphatic-to-aromatic ratio and PAH size in the local ISM \citep[e.g.,][]{Draine2021}. Recently, in a pilot study, \citet{Boersma2026} used available SPHEREx and Spitzer data from the PDR NGC\,2073 to study PAH emission variations across the region. They also measured how PAH sizes change when entering the more exposed and diffuse medium closer to the ionising star.  By correlating PAH spectral signatures with SKA-resolved AME intensity, spectral variations, and microwave emission morphology, a joint SPHEREx--SKA analysis will provide unique insight into how local physical conditions, such as radiation field strength, density, and grain charging, affect the efficiency of AME.

SPHEREx will measure the 3.3\,$\mu$m PAH emission feature in hundreds of thousands of galaxies across the sky, enabling population-level studies of nanograins and their associated radiation environments in massive galaxies of the nearby Universe. It will be capable of surveying PAH emission in relatively massive, star-forming systems out to $z\sim0.4$ \citep{Zhang2025}. At higher redshifts, the AME peak shifts toward SKA-Mid Band\,5b, enhancing the synergy between both facilities. SPHEREx will thus provide crucial measurements of the 3.3\,$\mu$m feature, enabling tests of whether variations in the small-grain population traced by PAHs are correlated with the strength and spectral properties of AME as constrained by SKA-Mid observations across different galaxy environments.

\paragraph{ALMA}

The Atacama Large Millimeter/submillimeter Array (ALMA), located on the 5000\,m Chajnantor plateau in Chile, consists of the 12-m main array with baselines up to 16\,km, complemented by the 7-m and 12-m antennas of the Atacama Compact Array that enhance surface-brightness sensitivity to extended emission. Its broad frequency range (35--950\,GHz) allows detailed characterisation of the Rayleigh--Jeans tail of thermal dust, providing crucial constraints on dust temperature, emissivity, and mass. ALMA Band~1 offers angular resolutions between 0.4$''$ and 31.8$''$ and covers the spectral region immediately above the peak of anomalous microwave emission (AME), making it well suited to isolate the transition between rotational and vibrational dust processes. The forthcoming Band~2 (67--116\,GHz) will further probe the overlap between spinning dust, free-free, and warm dust emission in environments where multiple mechanisms coexist. 

Although ALMA cannot access the low-frequency regime where the AME spectrum rises, its high-frequency sensitivity is essential for measuring the turnover of the spinning-dust spectrum and for disentangling the contribution from thermal dust. When combined with SKA-Mid observations, which anchor the low-frequency AME, ALMA enables continuous SED modelling over nearly three decades in frequency, yielding tight constraints on dust grain size distributions, charging, radiation field strength, and ISM conditions governing the production and efficiency of AME.

\section{Conclusions}
\label{sec:4-conclusions}

Anomalous microwave emission remains one of the least understood components of the
Galactic microwave sky. Although the spinning-dust hypothesis provides a compelling
explanation for most observations, the possible contribution from magnetic dust or other
nanoparticle populations cannot yet be excluded. The diversity of environments in which
AME is detected indicates that local physical conditions, such as radiation field strength,
density, and dust composition, contribute in shaping its properties.

High-resolution observations are crucial to constrain these factors. Instruments such as
the SKAO will provide the necessary combination of sensitivity and resolution to map AME
at arcsecond scales, revealing its spatial relation to photo-dissociation regions, \hii\
regions, and mid-infrared dust tracers. These data will test whether the spectral and
morphological variations of AME can be fully explained within the spinning-dust
framework or require additional emission mechanisms.

Polarisation measurements with SKA-Mid will also be decisive in discriminating between
spinning and magnetic dust, with upper limits at the percent level providing powerful
constraints on grain alignment and magnetic properties. In protoplanetary disks and
extragalactic environments, SKA-Mid observations will enable the separation of free–free,
synchrotron, and spinning dust components, helping to quantify the abundance and
evolution of very small grains.

Ultimately, AME observations with the SKAO will not only clarify the physical origin of
this enigmatic emission but also refine the modelling of foregrounds for next-generation
CMB experiments. The synergy between the SKAO, ALMA, and other observatories will
open a new observational window onto the microphysics of interstellar dust and the
processes that govern its evolution.


\newpage
\bibliographystyle{abbrvnat-maxbibnames4}
\bibliography{refs} 

\end{document}